\documentclass[12pt]{article}

\usepackage[dvips]{epsfig}
\usepackage[dvips]{graphics}

\def\be{\begin{equation}}
\def\ee{\end{equation}}

\newcommand{\black}[1]{\mathbf{#1}}
\newcommand{\base}[1]{\widehat{\black{e}}_{#1}}

\title{\bf `Circles in the Sky' in twisted cylinders}

\author{
G.I. Gomero\thanks{
german@ift.unesp.br}, \\
\\ 
Instituto de F\'{\i}sica Te\'orica, \\
Universidade Estadual Paulista,  \\
Rua Pamplona 145 \\
S\~ao Paulo, SP 01405--900, Brazil 
}

\begin{document}

\date{\today}

\maketitle

\begin{abstract} \noindent
It is shown here how prior estimates on the local shape of the universe can be 
used to reduce, to a small region, the full parameter space for the search of 
circles in the sky. This is the first step towards the development 
of efficient estrategies to look for these matched circles in order to detect a 
possible nontrivial topology of our Universe. It is shown how to calculate the 
unique point, in the parameter space, representing a pair of matched circles 
corresponding to a given isometry $g$ (and its inverse). As a consequence, (i) 
given some fine estimates of the covering group $\Gamma$ of the spatial section 
of our universe, it is possible to conf\/ine, in a very effective way, the 
region of the parameter space in which to perform the searches for matched 
circles, and reciprocally (ii) once identified such pairs of matched circles, 
one could determine with greater precision the topology of our Universe and our 
location within it.
\end{abstract}

It has recently been suggested that the quadrupole and octopole moments of the 
CMB anisotropies are almost aligned, i.e. each multipole has a 
preferred axis along which power is suppressed and both axes almost coincide. 
In fact, the angle between the preferred directions of these lowest multipoles 
is $\sim \! 10^\circ$, while the probability of this occurrence for two 
randomly oriented axes is roughly 1/62. There is also at present almost no 
doubt that the extremely low value of the CMB quadrupole is a real effect, i.e. 
it is not an illusion created by foregrounds \cite{TOCH}.

Traditionally, the low value of the quadrupole moment has been considered
as indirect evidence for a non--trivial topology of the universe. Actually, it 
was the fitting to these low values of the quadrupole and octopole moments of 
the CMB anisotropy which motivated the recent proposal that our Universe 
would be a Poincare's dodecahedron \cite{LWRLU}. On the other hand, the observed alignement of the quadrupole and the octopole moments
has recently been used as a hint for determining the direction along which might
occur the shortest closed geodesics characteristic of multiply connected
spaces \cite{OCTZH}. 

However, in most of the studies reported, the model
topology used for the comparison with data has been the $T^1$
topology, i.e. the torus topology with one scale of compactification of the
order of the horizon radius, and the other two much larger. This is the
simplest topology after the trivial one. Tests using $S$-statistics \cite{OCSS} and the \emph{circles in the sky}
method \cite{CSS} performed in \cite{OCTZH} yielded a null result for a
non--trivial topology of our universe. However it should be reminded that
multiply connected universe models cannot be ruled out on these grounds. In
fact, $S$-statistics is a method sensitive only to translational isometries,
while the search for the \emph{circles in the sky}, which in principle is
sensitive to detect any topology, was performed in a \emph{three-parameter 
version} able to detect translations only.

If the topology of the Universe is detectable in the sense of \cite{Detect},
then CMB anisotropy maps might present matched circles, i.e. pairs of circles
along of which the anisotropy patterns match \cite{CSS}. These circles are
actually the intersections (in the universal covering space of the spatial sections of spacetime) of the topological images of the sphere of last 
scattering, and hence are related by the isometries of the covering group 
$\Gamma$. Since matched circles will exist in CMB anisotropy maps of any 
universe with a detectable topology, i.e. regardless of its geometry and 
topology, it seems that the search for `circles in the sky' might be performed 
without any \emph{a priori} information of what the geometry and topology of 
the universe is. However, any pair of matched circles is described as a point 
in a six--dimensional parameter space, which makes a full--parameter 
search computationally expensive.%
\footnote{
These parameters are the center of each circle as a point in the
sphere of last scattering (four parameters), the angular radius of both
circles (one parameter), and the relative phase between them (one parameter).
}

Nevertheless, such a titanic search is currently being performed, and 
preliminary results have shown the lack of antipodal, and approximately 
antipodal, matched circles with radii larger than $25^\circ$ \cite{CSSK}. 
These results rule out the 
Poincare's dodecahedron model \cite{LWRLU}, and it has also been suggested 
that they rule out the possibility that we live in a small universe, since for 
the majority of detectable topologies we should expect antipodal or almost 
antipodal matched circles. In particular, it is argued that this claim is 
exact in all Euclidean manifolds with the only exception of the 
Hantzche--Wendt manifold ($\mathcal{G}_6$ in Wolf's notation \cite{Wolf}).

The purpose of this letter is twofold. First, it is shown how to use prior 
estimates on the local shape of the universe to reduce the region of the full 
parameter space in a way that the 
search for matched circles might become practical. In fact, it is shown how 
to calculate the unique point in the parameter space representing a pair of 
matched circles corresponding to a given isometry $g$ (and its inverse). As a 
consequence, given some fine estimates of the covering group $\Gamma$ of the 
present spatial section of our Universe, we may be able to confine, in a very 
effective way, the region of the parameter space in which to perform the 
searches for circles in the sky. This is the first important step towards the 
development of efficient estrategies to look for these matched circles. 
Moreover, once such pairs of matched circles had been identified, it is a 
simple 
matter to use its location in the parameter space to determine with greater 
precision the topology of our Universe.

Second, it emerges from the calculations that we should not expect (nearly) 
antipodal matched circles from the majority of detectable topologies. In 
particular, any Euclidean topology, with the exception of the torus, might 
generate pairs of circles that are not even nearly antipodal, provided the 
observer lies out of the axis of rotation of the isometry that gives rise 
to the pair of circles. This result might be generalized to the spherical case, 
for which work is in progress.

The main motivation for this work is the suspicion that the alignement of the 
quadrupole and the octopole moments of CMB anisotropies observed by the 
satellite WMAP, together with the \emph{anomalous} low value of the quadrupole 
moment, is the topological signature we should expect from a generic topology 
in a nearly flat universe, even if its size is slightly larger than the horizon 
radius. Moreover, as has been shown in \cite{LocSh}, if topology is detectable 
in a very nearly flat universe, the observable isometries will behave nearly as 
translations. If we locally approximate a nearly flat constant curvature space 
$M$ with Euclidean space, the smallest isometries of the covering group of 
$M$ will behave as isometries in Euclidean space. Since these isometries are 
not translations, they must behave as screw motions, thus an appropriate 
model to get a feeling of what to expect observationally in a nearly flat 
universe with detectable topology is a \emph{twisted} cylinder.

Thus, let us begin by briefly describing the geometry of twisted cylinders. An 
isometry in Euclidean 3-space can always be written as 
$(A,\black{a})$, where $\black{a}$ is a vector and $A$ is an orthogonal 
transformation, and its action on Euclidean space is given by 
\begin{equation}
\label{action}
(A,\black{a}) : \black{x} \mapsto A \black{x} + \black{a} \; ,
\end{equation}
for any point $\black{x}$. The generator of the covering group of a twisted 
cylinder is a screw motion, i.e. an isometry where its orthogonal part is a 
rotation and its translational part has a component parallel to the axis of 
rotation \cite{Wolf}. Thus we can always choose the origin and aligne the axis 
of rotation with the $z$--axis to write 
\be
\label{Rot}
A = \left( \begin{array}{ccc}
           \cos \alpha & - \sin \alpha & 0 \\
           \sin \alpha &   \cos \alpha & 0 \\
                0      &        0      & 1
           \end{array} \right)
\ee
for the orthogonal part, and
\be
\label{trans}
\black{a} = (0,0,L)
\ee
for the translational part of the generator $g = (A,\black{a})$. 

This is what is usually done when studying the mathematics of Euclidean 
manifolds, since it simplifies calculations. However, in cosmological 
applications this amounts to assume that the observer lies on the axis of 
rotation, which is a very unnatural assumption. In order to consider the 
arbitrariness of the position of the observer inside space, we parallel 
transport the axis of rotation, along the positive $x$--axis, a distance $\rho$ 
from the origin which remains to be the observer's position. Thus the 
generator of the twisted cylinder is now $g = (A,\black{b})$, with 
translational part given by 
\begin{equation}
\label{transGen}
\black{b} = \rho (1-\cos \alpha) \, \base{x} - \rho \sin \alpha \, \base{y} +
L \, \base{z} \; .
\end{equation}

The pair of matched circles related by the generator $g=(A, \black{b})$ are the 
intersections of the sphere of last scattering with its images under the 
isometries $g$ and $g^{-1}$ respectively, and the centers of these images are 
located at $g\black{0} = \black{b}$ and $g^{-1}\black{0} = -A^{-1}\black{b}$. Thus the angular positions of the centers of the matched circles are
\begin{equation}
\label{centmatcirc}
\black{n}_1 = \frac{\black{b}}{|\black{b}|} \qquad\mbox{and}\qquad
\black{n}_2 = - \, \frac{A^{-1}\black{b}}{|\black{b}|} = - A^{-1} \black{n}_1 
\; .
\end{equation}

There are four parameters we can determine using (\ref{Rot}) and 
(\ref{transGen}). The angle $\sigma$ between $\black{n}_1$ and the axis of 
rotation, the angle $\mu$ between the centers of the pair of matched circles, 
the angular size $\nu$ of both matched circles, and the phase--shift $\phi$. 
It turns out that only three of them are independent, as should be 
expected since a screw motion has only three free parameters $(\rho,L,\alpha)$.

We easily obtain
\begin{equation}
\label{angsep}
\cos \mu = \black{n}_1 \cdot \black{n}_2 = - \, \frac{1 + \tan^2 \sigma \cos
\alpha}{1 + \tan^2 \sigma}
\end{equation}
for the angular separation between both directions $\black{n}_1$ and 
$\black{n}_2$, while $\sigma$ is given by
\begin{equation}
\label{angb_rot}
\tan \sigma = \frac{\rho}{L} \, \sqrt{2(1- \cos \alpha)} \; .
\end{equation}
One can see from (\ref{angsep}) and (\ref{angb_rot}) that the matched circles 
will be antipodal only when the observer is on the axis of rotation ($\rho = 
0$), or the isometry is a translation ($\alpha = 0$). In particular, this 
shows that in a universe with any topology $\mathcal{G}_2$--$\mathcal{G}_6$, 
if a screw motion of the covering group generates a pair of matched circles, 
they will not necessarily appear nearly antipodal to an observer located off 
the axis of rotation. As an example consider an observer in a $\mathcal{G}_4$ 
universe located at a distance $\rho = L/2$ from the axis of rotation of 
the generator screw motion ($\alpha = \pi/2$). From (\ref{angsep}) and 
(\ref{angb_rot}) it follows that $\mu \approx 132^\circ$.

Next, to compute the angular size of these circles, let $R_{LSS}$ be the
radius of the sphere of last scattering. Simple geometry shows that, since
$|\black{b}|$ is the distance between the two centers of the spheres whose
intersections generate one of the matched circles, the angular size of this
intersection is
\begin{equation}
\label{angsize}
\cos \nu = \frac{|\black{b}|}{2R_{LSS}} = \frac{L}{2R_{LSS} \cos \sigma} \; .
\end{equation}

The computation of the phase--shift between the matched circles (the last
parameter we wish to constrain) is more involved. First we need to have an
operational definition of this quantity. This is simply accomplished if we
realize that there is a great circle that passes through the centers,
$\black{n}_1$ and $\black{n}_2$, of the matched circles. Orient this great
circle such that it passes first through $\black{n}_2$, and then through
$\black{n}_1$ along the shortest path, and let $\black{v}_2$, $\black{u}_2$,
$\black{u}_1$ and $\black{v}_1$ be the intersections of the great circle with
the matched ones following this orientation. If there were no phase--shift,
then we would have $g(R_{LSS}\black{u}_2) = R_{LSS}\black{u}_1$ and
$g(R_{LSS}\black{v}_2) = R_{LSS}\black{v}_1$. Hence we define the phase shift
as the rotation angle, around the normal of the sphere at $\black{n}_1$, that
takes $\black{u}_1$ to $\widehat{\black{u}}_2 =
g(R_{LSS}\black{u}_2)/R_{LSS}$, positive if the shift is counterclockwise,
and negative otherwise.

In order to use this operational definition to compute the phase--shift,
recall first that the great circle passing through $\black{n}_2$ and
$\black{n}_1$, with the required orientation, is given by
\begin{equation}
\label{greatcirc}
\black{n}(t) = \frac{1}{\sin \mu} \, [ \, \black{n}_2 \sin(\mu - t) +
\black{n}_1 \sin t \,] \; ,
\end{equation}
where $t$ is the angular distance between $\black{n}(t)$ and $\black{n}_2$.
Thus we have
\begin{eqnarray}
\label{prephase}
 \black{u}_1 & = & \frac{1}{\sin \mu} \, [ \, \black{n}_2 \sin \nu +
\black{n}_1 \sin (\mu - \nu) \,] \; , \nonumber \\
\black{u}_2 & = & \frac{1}{\sin \mu} \, [ \, \black{n}_2 \sin (\mu - \nu) +
\black{n}_1 \sin \nu \,] \; , \qquad\mbox{and} \\
\widehat{\black{u}}_2 & = & \frac{1}{\sin \mu} \, [ \, A\black{n}_2
\sin (\mu - \nu) + A\black{n}_1 \sin \nu \,] + \frac{\black{b}}{R_{LSS}} \; ,
\nonumber
\end{eqnarray}
Writing the positions of $\black{u}_1$ and $\widehat{\black{u}}_2$ with
respect to their projections to the axis $\black{n}_1$, as $\black{w}_1 = 
\black{u}_1 - \black{n}_1 \cos \nu$ and $\black{w}_2 = \widehat{\black{u}}_2 
- \black{n}_1 \cos \nu$, enables us to express easily the phase--shift as
\begin{equation}
\label{phase}
\cos \phi = \frac{\black{w}_1 \cdot \black{w}_2}{\sin^2 \nu} \; ,
\end{equation}
since $|\black{w}_1| = |\black{w}_2| = \sin \nu$. After a somewhat lengthy
calculation one arrives at
\begin{equation}
\label{postphase}
\cos \phi = \frac{2 (1 + \cos \alpha)}{1 - \cos \mu} - 1 
\end{equation}
for the phase--shift. It is easily seen that when the observer is on the axis 
of rotation ($\rho = 0$), the shift equals $\alpha$;%
\footnote{
We infere from this that, in the general case, both angles $\alpha$ and $\phi$ 
have the same sign.
} 
while when the isometry is a translation ($\alpha = 0$), the shift vanishes. In 
general, however, the shift depends on the three parameters $(\rho, L, 
\alpha)$, but only through the values of $\mu$ and $\alpha$, thus it is not an 
independent parameter. In fact, for a given pair $(\sigma, \mu)$, one can 
easily compute $\phi$, since $\alpha$ is readily obtained from (\ref{angsep}). 

Summarizing, given estimates of the parameters $(\rho, L, \alpha)$, and having 
determined an estimate of the axis of rotation of the screw motion, one can perform searches 
for pairs of circles both with centers at an angular distance $\sigma$ of this 
axis and separation $\mu$ between them, given by (\ref{angb_rot}) and 
(\ref{angsep}) respectively. The phase--shift between the circles is fixed by these two parameters. Moreover, 
we can also limit the search of matched circles to only those with angular size 
$\nu$ given by (\ref{angsize}). We have, thus, constrained three out of the six 
parameters needed to locate a pair of matched circles. The three missing 
parameters are the position ($\theta, \varphi$) of the axis of rotation of the 
screw motion and the azimuthal angle $\lambda$ of the center of one of the 
matched circles. 

Under the hypothesis that the alignement of the quadrupole and the octopole 
moments is due to the topology of space, the position of the axis of rotation 
might be estimated from this alignement. Work is in progress in this direction. 
Finally, only the angle $\lambda$ remains totally unconstrained.

Interestingly, a consequent precise identification of a pair of matched circles 
will allow, reciprocally, to determine with greater precision the same 
topological parameters with which we started, together with our position and 
orientation in the Universe.

\vspace{3mm}

\section*{Acknowledgments}

I would like to thank FAPESP for the grant under which this work was carried 
out (contract 02/12328--6). I also thank B. Mota, A. Bernui and W. 
Hip\'olito--Ricaldi for useful conversations.


%
%

\begin{thebibliography}{99}

\bibitem{TOCH} M. Tegmark, A. de Oliveira--Costa \& A.J.S. Hamilton, \emph{A
high resolution foreground cleaned CMB map from WMAP}, {\bf [astro-ph/0302496]}
(2003).

\bibitem{LWRLU} J-P. Luminet, J.R. Weeks, A. Riazuelo, R. Lehoucq \& J-P. Uzan, 
\emph{Nature} {\bf 425}, 593--595 (2003).

\bibitem{OCTZH} A. de Oliveira--Costa, M. Tegmark, M. Zaldarriaga \& A.J.S.
Hamilton, \emph{The significance of the largest scale CMB fluctuations in WMAP},
{\bf [astro-ph/0307282]} (2003).

\bibitem{OCSS} A. de Oliveira--Costa, G.F. Smoot \& A.A. Starobinsky, \emph{Ap.
J.} {\bf 468}, 457--461 (1996); Proc. from XXXIst Recontres de Moriond: Future 
CMB missions, {\bf [astro-ph/9705125]}.

\bibitem{CSS} N.J. Cornish, D. Spergel \& G. Starkman, \emph{Class. Quantum
Grav.}, {\bf 15}, 2657--2670 (1998).

\bibitem{Detect} G.I. Gomero, M.J. Rebou\c{c}as \& R. Tavakol, \emph{Class.
Quantum Grav.} {\bf 18}, 4461--4476 (2001); {\bf 18}, L145--L150 (2001); 
\emph{Int. J. Mod. Phys. A} {\bf 17}, 4261--4272 (2002). \\
E. Gausmann, R. Lehoucq, J.-P. Luminet, J.-P. Uzan \& J. Weeks, \emph{Class.
Quantum Grav.} {\bf 18}, 5155-- (2001). \\
J.R. Weeks, \emph{Detecting topology in a nearly flat hyperbolic universe},
{\bf [astro-ph/0212006]} (2002). \\
G.I. Gomero \& M.J. Rebou\c{c}as, \emph{Phys. Lett. A} {\bf 311}, 319--330
(2003).

\bibitem{CSSK} N.J. Cornish, D. Spergel, G. Starkman \& E. Komatsu, 
\emph{Constraining the Topology of the Universe}, {\bf [astro-ph/0310233]}, 
submitted to PRL.

\bibitem{Wolf} J. Wolf, \emph{Spaces of Constant Curvature}, 5th edition, 
Publish or Perish, Houston (1984).

\bibitem{LocSh} B. Mota, G.I. Gomero, M.J. Rebou\c{c}as \& R. Tavakol, 
\emph{What do very nearly flat detectable cosmic topologies look like?}, 
{\bf [astro-ph/0309371]}, submitted to PRL.

\end{thebibliography}
\end{document}